# Pre-perihelion observations of comet P/2023 JN16 (Lemmon) at the 2024 apparition

Serhii Borysenko[1], Alexander Baransky[2], Christopher A. Onken[3], Iryna Kulyk[1]

[1]Main Astronomical Observatory of NAS of Ukraine, Akademika Zabolotnoho 27, Kyiv 03143, Ukraine

e-mail: borisenk@mao.kiev.ua

[2]Astronomical Observatory of Taras Shevchenko National University of Kyiv Observatorna str. 3, Kyiv, 04053 Ukraine

[3]Research School of Astronomy and Astrophysics, Australian National University, Canberra, ACT 2611, Australia

**ABSTRACT**

We present an analysis of the photometric data of new main belt comet P/2023 JN16 (Lemmon) observed with the 1.35-m SkyMapper telescope at Siding Spring Observatory in 2024 during July – September. The faint object showed dusty activity during a long period. Some photometric parameters, such as SkyMapper $r$ band magnitudes, $Af\rho$ parameters, and absolute magnitude were estimated. We estimate the nuclear radius for comet P/2023 JN16 to be less than 0.5 km. The activity of the object is possible driven by the release of large, slow-moving particles.

1. Introduction

Cometary activity has been observed for minor solar system bodies in a wide range of heliocentric distances. The observed activity patterns can serve as an indicator of the volatile distribution over the Solar system (Meech & Svoren, 2004). Recently, the active bodies exhibiting cometary comae and tails were found in the innermost region of the Solar system among the "classical" asteroids of the Main Asteroid Belt. The subgroup of comet-like bodies on asteroidal orbits was recognized as Active Asteroids or Main-Belt Comets (MBCs) (Hsieh& Jewitt, 2006; Jewitt et al., 2015). These authors give the working definition for this population: active asteroids are small bodies that: 1) have semimajor axis $a$ less than Jupiter's semimajor axis, 2) Tisserand parameter $Tj > 3.08$ (for definition of Tisserand parameter see (Kosai, 1992), 3) exhibit resolved coma or tail indicating the mass-loss.

Main-Belt Comet P/2023 JN16 (Lemmon) presents a particularly interesting case. The Horizon system classifies it as an object belonging to the Encke-comet dynamical group[1], however, its Tisserand parameters ($T_{Jup}$ = 3.351) exceeds that of the comet 2P/Encke group, and following the definition noted above, P/2023 JN16 could belong to the MBC family. The orbital parameters of P/2023 JN16 are presented in Table 1.

Initially discovered as an asteroid on May 5, 2023 by the Mt. Lemmon Survey, this object exhibited no signs of cometary activity during its early observations[2]. However, in June 2023, K. Ly (El Monte, CA, USA) and Arnt Schnabel (Leipzig, Germany) found a slightly elongated image with a straight 2" comet-like tail (Deen, S. et al., 2024). The next observations on July 15, 2023 showed a very condensed 6" diameter coma (V magnitude was between 20.7 and 20.8) with no tail (Ly, K., et al. 2024). The follow-up observations made on September 5, 2023 with the 4.3-m Lowell Discovery Telescope revealed a faint 15" long tail. The additional images taken by Arnt Schnabel and D. Ruhela on June 29, 2024 showed this object to have an odd appearance: "obviously highly diffuse" with no apparent condensation (V magnitude is 22.0) and with a tail at least of 60" long (Lyet al., 2024). Following an alert that 2023 JN16 can be a newly activated Main Belt Asteroid, QZ Ye and M. S. P.Kelley (University of Maryland) and D. Bodewits (Auburn University), on behalf of the Zwicky Transient Facility (ZTF) Partnership, examined the ZTF images obtained in 2023 for pre-discovery detections. The only pre-discovery image was found in ZTF data gathered on April 29, 2023, in which the object had r-magnitude of 19.1 (PS1 photometric system) (Ly et al., 2024).

---

[1]https://ssd.jpl.nasa.gov/tools/sbdb_lookup.html#/?sstr=P%2F2023%20JN16
[2]https://minorplanetcenter.net/iau/ECS/MPCArchive/2023/MPS_20230615.pdf

The present paper discusses a series of pre-perihelion photometric observations obtained for comet P/2023 JN16 (Lemmon) during its 2024 apparition using the SkyMapper telescope at Siding Spring Observatory (MPC code – Q55). The observations are described in Sect.2, data analysis is discussed in Sect.3, the analysis and interpretation is presented in Sect. 4.

## 2. Observations

Observations were conducted between 29 July, 2024 and 5 September, 2024 with the SkyMapper telescope. Table 2 gives the observing log: data of the observations, number of 60-sec exposures, heliocentric and geocentric distances of the comet during the observations, phase angles and true anomaly angles. The SkyMapper has an f/4.8 focal ratio, making it highly efficient as a survey instrument (Onken et al., 2024). A mosaic of 16 CCDs, each having 2048×4096 pixels with a scale of ~ 0.5 arcsec was used as a detector[3]. Each CCD has a field of view of 34' × 17' resulting in a whole field of view of 2.4×1.2 deg$^2$. The typical readout noise is about 10 $e^-$ and the conversion gain is about 1 $e^-$/ADU.

---

[3] The full SkyMapper mosaic has 32 CCDs, but only half of the array was operational at the time of observation.

The filter set is broadly similar to the SDSS photometric system, with some important differences (Bessel et al. 2011). For the observations we utilized *g, r, i*-filters, however, taking into account the faint appearance of P/2023 JN16 during the observation period, in this report we present only *r*-band photometry. The 50% cut-on and cut-off edges of the *r*-filter are 5385Å and 6965Å, respectively, and FWHM is 1580Å (Bessel et al., 2011). All the images were pre-processed with the SkyMapper Science Data Pipeline which accounts for bias, flatfield, and amplifier additive offsets, as well as generating a World Coordinate System solution (for details, see Onken et al., 2024).

[Fig 1]

[Table 1]

### 3. Data analysis

The observations were conducted in July – September 2024 during 12 nights. From the full dataset we selected 8 nights (see Table 2) when the comet was far from bright stars, and it was possible to make accurate measurements. We used the SkyMapper Sourthern Survey (SMSS) DR4 as a reference photometric catalogue (Onken et al., 2024). Each CCD (34' × 17') contains dozens of SMSS reference photometric stars, but we select only those for which the deviation from the Sun color indices do not exceed $0.02^m - 0.03^m$. Thus, on each frame we had 6 – 12 stars with the color indices close to the Sun, namely, $g - r = 0.28$ mag and $r - i = 0.14$ mag (Willmer, 2018; Wolf et al., 2018).

Free software *Aperture Photometry Tools* was used as an instrument for aperture photometry (Laher et al., 2012). Firstly, we planned to measure the total image of the comet with an elliptical aperture, but due to the low signal-to-noise ratio and possible high photometric errors, we used only the head of the comet for photometry with a circular aperture having a radius about of 1×FWHM depending on seeing.

### 4. Results and discussion

The comet passed opposition in late August (Table 1) and changed its direction relative to the Sun (Fig. 1). The anti-motion vector *(-v)* showed low changes throughout the entire observational period. Faint, short and thin dust tail oriented along the *-v* direction was visible during all the nights. Fig 2. presents dependence of the visible magnitude measured within the aperture chosen on the solar phase angle. The brightening of the near-nuclear area measured within the aperture is probably caused by the phase (oppositional) effect (Liang et al., 2023).

The *Afρ* parameter was calculated from the circular aperture with the expression (A'Hearn et al., 1984):

$$Af\rho = 4r^2\Delta^2 \times 10^{0.4(m_{Sun} - m_c)}\rho^{-1} \quad (1)$$

here *r* (AU) is the heliocentric distance; $\Delta$ (cm) is the geocentric distance; $m_{Sun}$, $m_c$ are the apparent SkyMapper *r* magnitudes of the Sun ($m_{Sun}$ = -26.91; Willmer, 2018) and of the comet, respectively; $\rho$[cm] is the radius of the photometric aperture projected onto the sky. The calculated values of the *Afρ* parameter are low and vary from 0.91 ± 0.08 cm to 1.56 ± 0.16 cm. The maximum value of the *Afρ* was obtained for the observations made on 23 August when the phase angle had a minimal value of 3.4°. The similar small values of *Afρ* (measured in broadband filters of Johnson-Cousins photometric system) were obtained for main belt comets and active asteroids (133P/Elst – Pizarro, 238P/Read, 288P/300163, 311P/PANSTARRS, 331P/Gibbs) (Borysenko et al., 2020a). The unusually weak dust activity of P/2023 JN16, is possibly related to the predominance of large dust grains and the small size of its nucleus. In visible light, the brightness of a coma is more sensitive to the optical scattering area rather than the dust mass. Per unit mass, small particles provide a much larger area than large ones. Therefore, with a coarse-grained mixture, the optical brightness (and *Afρ*) drops, even if the dust mass emission is considerable (Haslebacher et al., 2024; Kereszturi et al., 2025).

A rough estimation of the upper limit of a radius for the cometary nucleus was calculated as (Russel, 1916):

$$R^2_n = 2.238 \times 10^{22} r^2 \Delta^2 10^{0.4(m_\odot - m_{nucl} + \alpha\beta)} A^{-1} \quad (2)$$

here $r$ [AU] is the heliocentric distance; $\Delta$ [AU] is the geocentric distance; $m_\odot$ and $m_{nucl}$ are the apparent SkyMapper $r$ magnitude of the Sun and an estimated nucleus magnitude, respectively; $\alpha$ is the phase angle and $\beta$ = 0.04 mag/deg – phase coefficient (Lamy et al., 2004). The geometric albedo value A = 0.05 was used (Hsieh et al., 2009). A rough estimation of radius gives a value for nuclear radius $R_n$ ≤ 0.47 ± 0.02 km (Table 3). This value is close to the nucleus estimates for some active main belt objects, such as 238P/Read, 311P/PANSTARRS, P/2010 R2 (La Sagra) (Hsieh et al., 2011; Jewitt et al., 2013; Jewitt et al., 2014; Hsieh et al., 2014) and for active asteroid P/2017 S5 (ATLAS) (Jewitt et al., 2019; Borysenko et al., 2019).

We model the comet tail with the Finson – Probstein technique (Vincent, 2014). The model results show distribution of dust along a synchrone with t = 300 days. The integration step was 1 day. The model parameter $\beta = 0.57 Q_{pr}/\rho a$, where $\rho$ is the density of the dust grain, expressed in g/cm$^3$, $a$ is the radius of the dust grain expressed in micrometers, $Q_{pr}$ is the efficiency of the radiation pressure, which depends on the size, shape and optical characteristics of a dust grain (for cometary dust, the radiation pressure efficiency is adopted to be about 1). Then, taking the density of cometary dust for the main-belt comets to be about 1 g/cm$^3$ (Moreno et al., 2012), we can obtain information about sizes of the particles in the coma and tail (Kelley et al., 2013). A range of $\beta$ values are 0.0004, 0.0006, and 0.0008. In accordance with modeling, the large particles with radius of $a$ >700 μm dominate the inner regions of the coma for comet P/2023 JN16 (Fig. 2). Such large values are in good agreement with the very low calculated $Af\rho$ parameters for the comet because large grains give a smaller scattering area per unit of mass. Compared to active asteroids with stronger dust signatures (e.g. Gault; Borysenko et al., 2020b), P/2023 JN16 exhibits an $Af\rho$ an order of magnitude lower, consistent with extremely weak dust production.

**[Fig 2]**

The observed characteristics support the interpretation that the activity is driven by the release of large, slow-moving particles from a small, low-gravity nucleus. These findings contribute to the growing evidence for the diversity of activity patterns among main-belt comets and provide new constraints on dust emission processes in small, weakly active bodies. The emergence of the tail without a visible nucleus suggests that the object's activity is dominated by sublimation processes affecting only its outer layers, while the nucleus remained obscured or too faint to detect.

**Acknowledgements**

This work was funded by Grant of the Ukraine – Australia Research Fund to realize project "Distantly active objects in the Solar system: variety of the activity patterns and physical properties of dust comae". The Australian Academy of Science has partnered with the Breakthrough Prize Foundation to deliver a program to support Ukrainian researchers in eligible fields of science who have been impacted by the war with Russia. The donation is being used to fund activities with practical support to enable the continuation of research and technology activities by Ukrainian scientists.

We thank the Australian National University, director of the Siding Spring Observatory Christian Wolf, director of Programs of Astronomy Australia Ltd. Dr

James Murray. Special thank Fabio Caviccio (MSB Software) for the provided licensed *Astroart 9* software.

This research has made use of the scientific software at www.comet-toolbox.com

Table 1. P/2023 JN16 (Lemmon) orbital parameters

| $q^a$, AU | $e^b$ | $a^c$, AU | $i^d$,(°) | $Q^e$, AU | UT, perihelion passage | $P^f$, year | $T^g_{jup}$ |
|---|---|---|---|---|---|---|---|
| 2.300 | 0.147 | 2.697 | 3.703 | 3.092 | 2024-Dec-29.948 | 4.428 | 3.351 |

[a]Perihelion distance
[b]Eccentricity
[c]Semimajor axis
[d]Inclination of orbit
[e]Aphelion distance
[f]Orbital period.
[g]Tisserand parameter.

Table 2 Log of observations

| Date | $N_i \times Exp^a$., s | Median seeing, " | $r^b$, AU | $\Delta^c$, AU | $\alpha^d$, deg | $\nu^e$, deg |
|---|---|---|---|---|---|---|
| July, 30 | 49 x 60 | 3.3 | 2.390 | 1.417 | 9.1 | 314.934 |
| August, 3 | 46 x 60 | 2.4 | 2.385 | 1.398 | 7.3 | 316.058 |
| August, 6 | 24 x 60 | 2.7 | 2.382 | 1.385 | 5.9 | 316.922 |
| August, 7 | 48 x 60 | 2.5 | 2.381 | 1.382 | 5.5 | 317.191 |
| August, 23 | 25 x 60 | 2.5 | 2.365 | 1.359 | 3.4 | 321.717 |
| September, 2 | 25 x 60 | 3.3 | 2.356 | 1.378 | 7.9 | 324.562 |
| September, 3 | 24 x 60 | 2.1 | 2.355 | 1.381 | 8.4 | 324.848 |
| September, 5 | 26 x 60 | 2.4 | 2.353 | 1.388 | 9.3 | 325.426 |

[a] Number of stacking images and exposure time of each;
[b] Heliocentric distance;
[c] Geocentric distance;
[d] Phase angle;
[e] True anomaly

Table 3 Results of measurements

| Date | $r^a$ | $Af\rho^b$, cm | $R_n^c$, km |
|------|-------|----------------|-------------|
| July, 30     | 22.11 ± 0.08 | 0.99 ± 0.07 | 0.42 ± 0.02 |
| August, 3    | 21.89 ± 0.08 | 1.19 ± 0.09 | 0.44 ± 0.02 |
| August, 6    | 21.89 ± 0.10 | 1.18 ± 0.11 | 0.43 ± 0.02 |
| August, 7    | 22.16 ± 0.10 | 0.91 ± 0.08 | 0.37 ± 0.02 |
| August, 23   | 21.55 ± 0.11 | 1.56 ± 0.16 | 0.47 ± 0.03 |
| September, 2 | 22.03 ± 0.11 | 1.01 ± 0.10 | 0.41 ± 0.02 |
| September, 3 | 21.93 ± 0.09 | 1.11 ± 0.09 | 0.43 ± 0.02 |
| September, 5 | 21.96 ± 0.10 | 1.08 ± 0.10 | 0.44 ± 0.02 |

[a] Apparent SkyMapper *r* magnitude of the comet
[b] $Af\rho$ parameter of the comet;
[c] Upper limit of radius for the cometary nucleus;

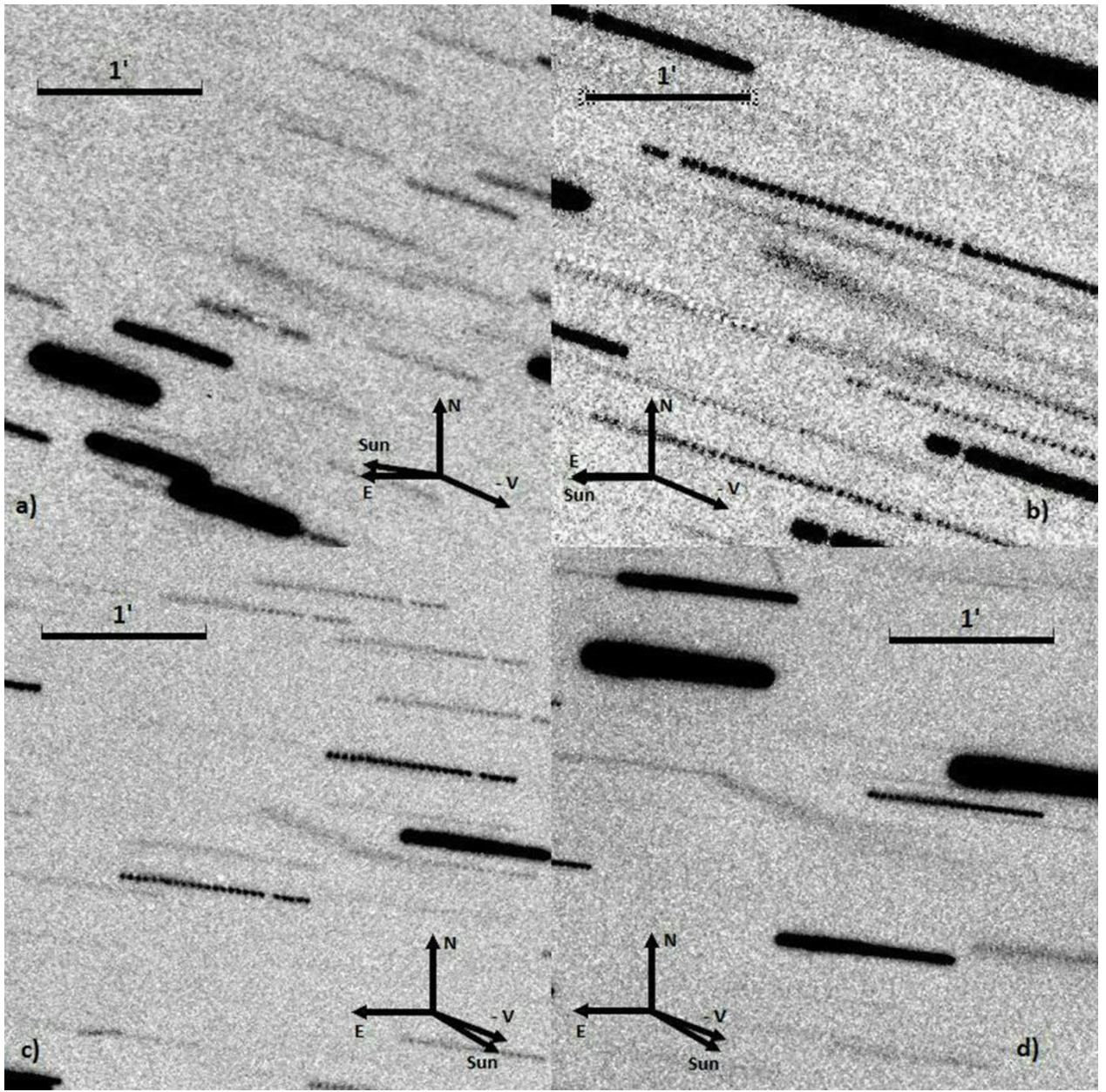

**Fig 1**

Samples of the obtained images for comet P/2023 JN16 Lemmon: a)2024-07-30, b) 2024-08-07), c) 2024-09-03, d)2024-09-05.

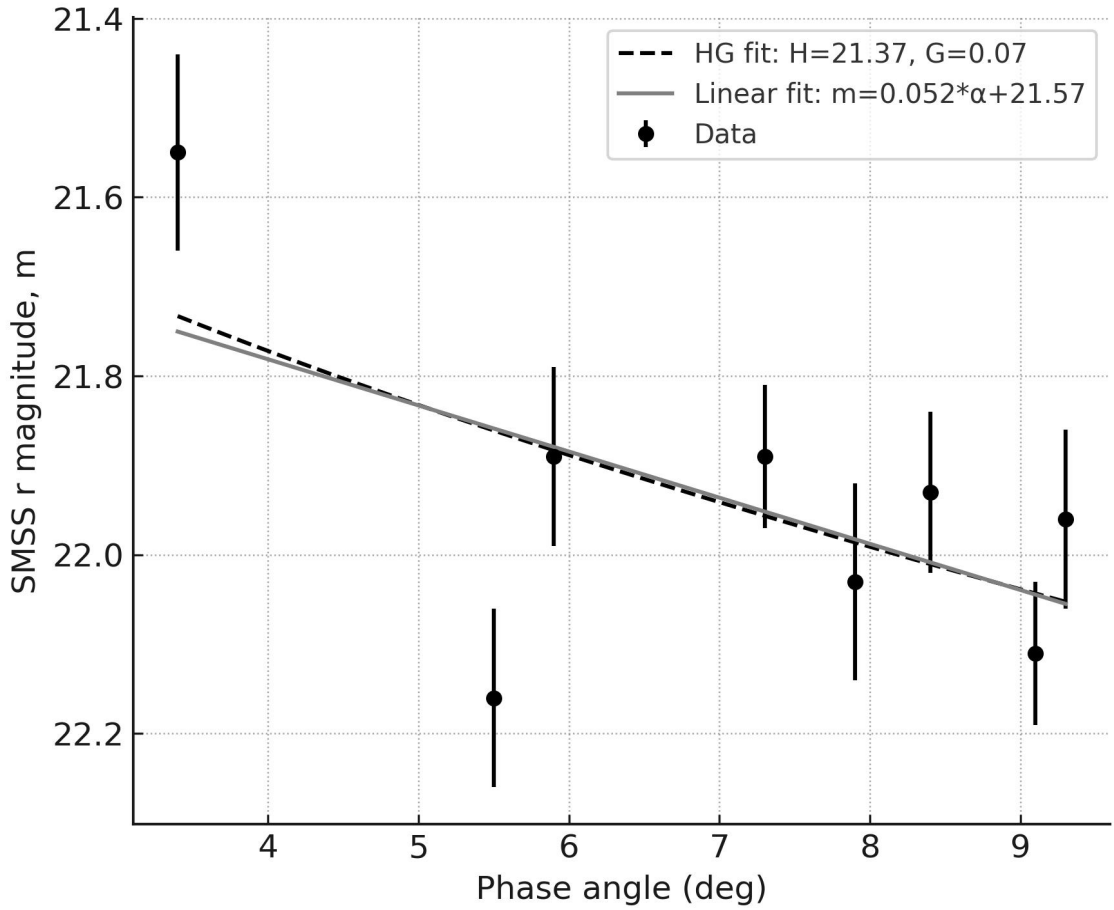

**Fig 2.** The results of approximation by Bowel et al., 1989 (dashed line), where m(α) = H – 2.5log$_{10}$[(1 – G)Φ$_1$(α) + GΦ$_2$(α)] and linear approximation (solid line), where m(α) = m$_0$ + kα.

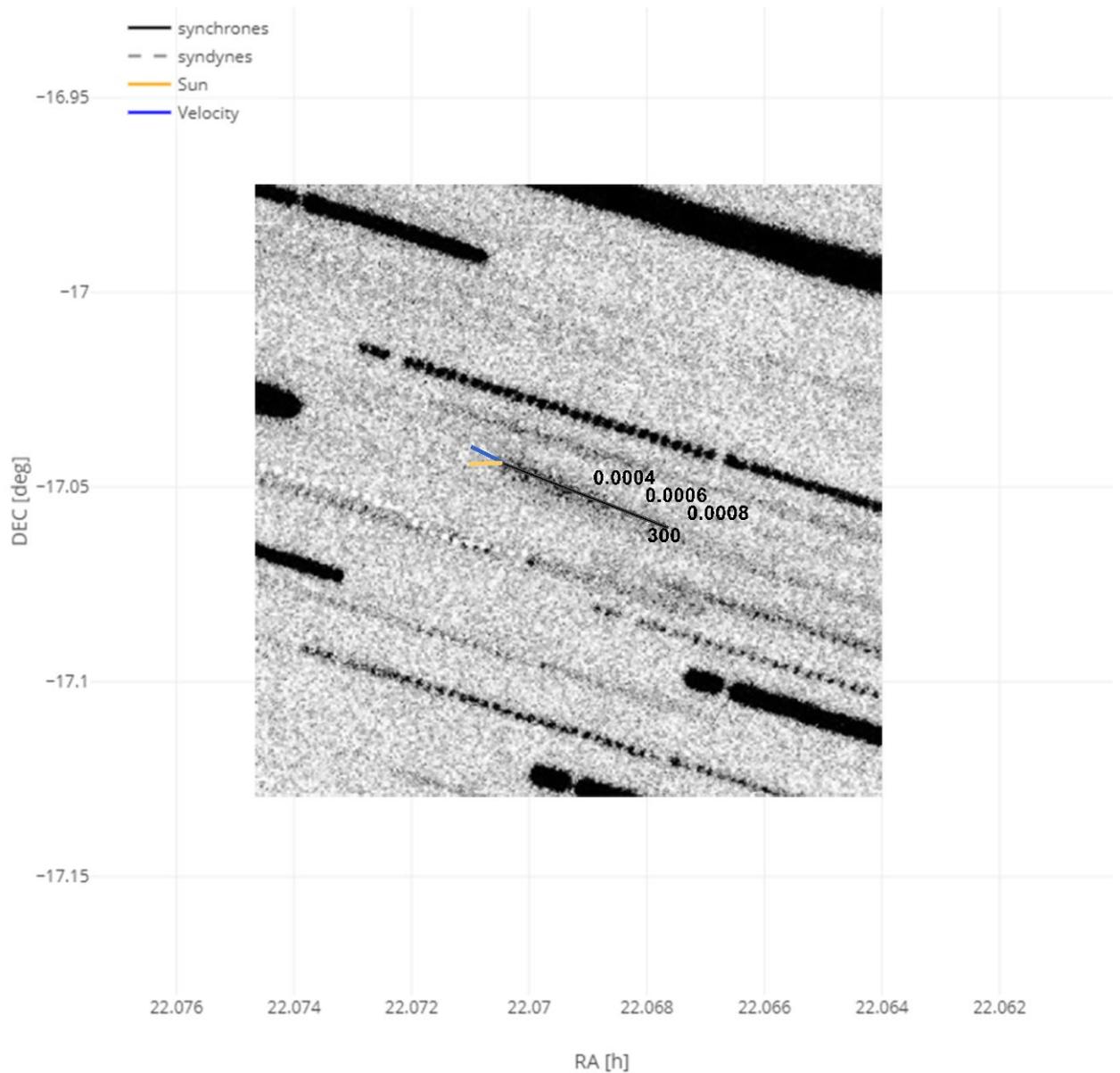

**Fig 3**

Finston – Probstein diagram of comet P/2023 JN16 (Lemmon) superimposed on a grid in real scale for original *r*-band image of the comet obtained in 2024 (August 7, total expos. 48 × 60 s) with 1.35 - m SkyMapper telescope of the Siding Spring Observatory (Australia), showing the distribution of beta-parameters along a synchrone with t = 300 days (Vincent, 2014).